\documentclass[a4paper,twoside]{article}

\usepackage{ams2la,fancyhdr,multicol,graphicx,bm,ccmap}
\usepackage[dvipdfm, a4paper, pdfstartview=FitH, bookmarks=false,
        colorlinks=false,
        linkcolor=blue, urlcolor=blue, pdfborder=001, citecolor=blue,
        pdftitle={Changepdftitle}, pdfauthor={Changepdfauthor},
        pdfcreator=CHIN.\ PHYS.\ LETT., pdfproducer=CPL,
        pdfkeywords={changePACS}]{hyperref}

\def\headrule{\kern 1mm \hrule width 17cm \kern -1mm}%
\def\footnoterule{\kern 1mm \hrule width 7cm \kern 2.2mm}%
\def\REF#1{\par\hangindent\parindent\indent\llap{#1\enspace}\ignorespaces}%

\newcommand{\cplyear}{} \newcommand{\cplvol}{}
\newcommand{\cplno}{} \newcommand{\cplpagenumber}{}
 \newcommand{\cplpage}{\cplpagenumber-\thepage}
\begin{document}

\begin{center}

\large\bf{\boldmath{Hadronic decays of the spin-singlet heavy quarkomium under the principle of maximum conformality}} \footnote{supported by the Fundamental Research Funds for the Central Universities under Grant No.CQDXWL-2012-Z002, the Program for New Century Excellent Talents in University under Grant No.NCET-10-0882, and the Natural Science Foundation of China under Grant No.11275280 and No.11375279.

\hspace{1.8mm}$^{**}$ Correspondence author. Email: wuxg@cqu.edu.cn

\hspace*{1.8mm}\copyright\,{\cplyear}
\href{http://www.cps-net.org.cn}{Chinese Physical Society} and
\href{http://www.iop.org}{IOP Publishing Ltd}}
\\[4mm]
\normalsize\rm{}Zhang Qiong-Lian, Wu Xing-Gang$^{**}$, Zheng Xu-Chang, \\
Wang Sheng-Quan, Fu Hai-Bing, Fang Zhen-Yun
\\[1mm]\small\sl Department of Physics, Chongqing University, Chongqing 401331

\end{center}

\vskip -1mm

\noindent{\narrower\small\sl{}

The principle of maximum conformality (PMC) provides a way to eliminate the conventional renormalization scale ambiguity in a systematic way. By applying the PMC scale setting, all non-conformal terms in perturbative series are summed into the running coupling, and one obtains a unique, scale-fixed prediction at any finite order. In the paper, we make a detailed PMC analysis for the spin-singlet heavy quarkoniums decay (into light hadrons) at the next-to-leading order. After applying the PMC scale setting, the decay widths for all those cases are almost independent of the initial renormalization scales. The PMC scales for $\eta_c$ and $h_c$ decays are below $1$ GeV, in order to achieve a confidential pQCD estimation, we adopt several low-energy running coupling models to do the estimation. By taking the MPT model, we obtain: $\Gamma(\eta_{c} \to LH)=25.09^{+5.52}_{-4.28}$ MeV, $\Gamma(\eta_{b} \to LH)=14.34^{+0.92}_{-0.84}$ MeV, $\Gamma(h_{c} \to LH)=0.54^{+0.06}_{-0.04}$ MeV and $\Gamma(h_{b} \to LH)=39.89^{+0.28}_{-0.46}$ KeV, where the errors are calculated by taking $m_{c}\in[1.40\rm GeV,1.60\rm GeV]$ and $m_{b}\in[4.50\rm GeV,4.70\rm GeV]$. These decay widths agree with the principle of minimum sensitivity estimations, in which the decay widths of $\eta_{c,b}$ are also consistent with the measured ones.

\par}
\vskip 3mm\normalsize

\noindent{\narrower\sl{PACS: 12.38.Bx, 14.40.Pq, 13.25.Ft}
{\rm\hspace*{13mm}DOI:}

\par}\vskip 3mm

\begin{multicols}{2}

Heavy quarkonium plays an important role in understanding the QCD factorization theory. Its inclusive annihilation provides one of the most interesting topics in heavy quarkonium physics. The annihilation of the spin-singlet $S$- and $P$-wave heavy quarkonium states to light hadrons (LHs) have been studied at the order of ${\cal O}(\alpha_{s}\upsilon^{2})$$^{[1,2]}$ within the framework of nonrelativistic QCD (NRQCD)$^{[3]}$. However, those estimations suffer from large renormalization scale uncertainties. To improve the accuracy of pQCD estimations, we adopt the principle of maximum conformality (PMC)$^{[4-10]}$ to set the renormalization scale.

To apply PMC, one can first finish the renormalization procedure by using an arbitrary initial scale $\mu_{R}^{\rm init}$ (in the perturbative region), and set the optimal (PMC) scales by absorbing all non-conformal terms into the running coupling via a step-by-step way. The decay widths for $|H[n]\rangle\to LH$ up to ${\cal O}(\alpha_s v^2)$ can be written as :
\begin{equation}
\Gamma(H[n]) = C_{0}([n])\alpha_{s}^{2}(\mu_{R}^{\rm init})\left[1+\frac{\alpha_{s}(\mu_{R}^{\rm init})} {\pi}C_{1}([n])\right],
\end{equation}
where $H$ stands for charmonium or bottomonium, and $[n]$ stands for the color-singlet state $^{1}S^{[1]}_{0}$ or $^{1}P^{[1]}_{1}$ respectively. The leading-order (LO) coefficients
\begin{eqnarray}
&&\!\!\!\!\!\!\!\!\!\!\!\!\!\! C_{0}([^{1}S^{[1]}_{0}]) = \frac{4\pi}{9m_{Q}^{2}}\left[\langle \mathcal{O}(^{1}S_{0}^{[1]})\rangle_{^{1}S_{0}}-\frac{4\langle \mathcal{P}(^{1}S_{0}^{[1]})\rangle_{^{1}S_{0}}}{3m_{Q}^{2}}\right], \nonumber
\end{eqnarray}
\begin{eqnarray}
&&\!\!\!\!\!\!\!\!\!\!\!\!\!\! C_{0}([^{1}P^{[1]}_{1}]) = \frac{5\pi}{6m_{Q}^{2}}\left[\langle \mathcal{O}(^{1}S_{0}^{[8]})\rangle_{^{1}P_{1}}-\frac{4\langle \mathcal{P}(^{1}S_{0}^{[8]})\rangle_{^{1}P_{1}}}{3m_{Q}^{2}}\right], \nonumber
\end{eqnarray}
where $\langle\mathcal{O}\rangle$ and $\langle\mathcal{P}\rangle$ are long-distance matrix elements (LDMEs), the superscript $[1]$ or $[8]$ stands for the color-singlet or color-octet state, respectively. The NLO coefficient $C_{1}([n])$ can be divided into $\beta$-dependent non-conformal and $\beta$-independent conformal parts,
\begin{equation}
C_{1}([n]) = C_{1}^{(\beta)}([n])\beta_{0}+C_{1}^{\rm (con)}([n]), \nonumber
\end{equation}
where $\beta_{0}=11-2n_{f}/3$. For the conventional scale setting, the scale is fixed once it has been set to an initial value, i.e. one usually takes $\mu_{R}\equiv \mu_R^{\rm init}=2m_Q$ for estimating the heavy quarkonium decays. After applying the PMC scale setting, the $\beta_0$-dependent terms can be absorbed into PMC scale and we obtain
\begin{displaymath}
\Gamma(H[n])=C_{0}([n])\alpha_{s}^{2}(\mu_{R}^{\rm PMC})\left[1+\frac{\alpha_{s}(\mu_{R}^{\rm PMC})}{\pi}C_{1}^{\rm (con)}([n])\right],
\end{displaymath}
where $\mu_{R}^{\rm PMC}=\mu_{R}^{\rm init} \exp\left[-C_{1}^{(\beta)}([n])\right]$. The non-conformal $C_{1}^{(\beta)}([^{1}S_{0}])$ and $C_{1}^{(\beta)}([^{1}P_{1}])$, and the conformal $C_{1}^{\rm (con)}([^{1}S_{0}])$ and $C_{1}^{\rm (con)}([^{1}P_{1}])$ at the initial scale $\mu_{R}^{\rm init}$ can be derived from Refs.[1,2], i.e.
\begin{eqnarray}
&& \!\!\!\!\!\!\!\!\!\!\! C_{1}^{(\beta)}([^{1}S_{0}]) = -\frac{a}{a+b}\frac{1}{72} \left[36\ln\left(\frac{4m_{Q}^{2}} {\left(\mu^{\rm init}_{R}\right)^{2}} \right)-96\right] \nonumber \\
&&\quad\quad\quad\quad -\frac{b}{a+b}\frac{1}{144}\left[72\ln\left(\frac{4m_{Q}^{2}}{\left(\mu^{\rm init}_{R}\right)^{2}}\right) -246\right] \nonumber \\
&&\!\!\!\!\!\!\!\!\!\!\!  C_{1}^{\rm (con)}([^{1}S_{0}]) = -\frac{a}{a+b}\frac{1} {72}(93\pi^{2}-852)- \frac{b}{a+b}\frac{1}{144}\times \nonumber \\
&&\quad\quad\quad\quad\quad \left[192\ln\left(\frac{\mu_{F}^{2}}{4m_{Q}^{2}}\right) + 237\pi^{2}-2258\right], \nonumber
\end{eqnarray}

\begin{eqnarray}
&& \!\!\!\!\!\!\!\!\!\!\! C_{1}^{\rm (con)}([^{1}P_{1}]) =-\frac{A}{A+B} \frac{1}{72}(129\pi^{2}-1392) \nonumber \\
&&-\frac{B}{A+B} \frac{1}{288}\left[168\ln\left(\frac{\mu_{F}
^{2}}{4m_{Q}^{2}}\right)+735\pi^{2}-6892\right] \nonumber \\
&& +\frac{C}{A+B}\left[7\pi^{2}-112-24\ln\left(\frac{\mu_{F}^{2}} {4m_{Q}^{2}}\right)\right]  \nonumber\\
&& +\frac{D}{A+B}\left[1740\ln\left(\frac{\mu_{F}^{2}}
{4m_{Q}^{2}}\right)-555\pi^{2} +9236\right], \nonumber \\
&& \!\!\!\!\!\!\!\!\!\!\! C_{1}^{(\beta)}([^{1}P_{1}]) =-\frac{A}{A+B}\frac{1}{72}\left[36\ln \left(\frac{4m_{Q}^{2}}{\left(\mu^{\rm init}_{R}\right)^{2}}\right)-96\right]- \nonumber \\
&&\frac{B}{A+B}\frac{1}{288}\left[144\ln \left(\frac{4m_{Q}^{2}}{\left(\mu^{\rm init}_{R}\right)^{2}}\right)-492\right], \nonumber
\end{eqnarray}
where $\mu_F$ is the factorization scale and the coefficients
\begin{eqnarray}
&& \!\!\!\!\!\!\!\!\!\!\!  a=\frac{4\pi}{9m_{Q}^{2}}\langle \mathcal{O}(^{1}S_{0}^{[1]})\rangle_{^{1}S_{0}},
b=-\frac{16\pi}{27m_{Q}^{4}}\langle \mathcal{P}(^{1}S_{0}^{[1]}) \rangle_{^{1}S_{0}}, \nonumber \\
&& \!\!\!\!\!\!\!\!\!\!\!  A=\frac{5\pi}{6m_{Q}^{2}}\langle \mathcal{O}(^{1}S_{0}^{[8]})\rangle_{^{1}P_{1}},
B=-\frac{10\pi}{9m_{Q}^{4}}\langle \mathcal{P}(^{1}S_{0}^{[8]})\rangle_{^{1}P_{1}}, \nonumber \\
&& \!\!\!\!\!\!\!\!\!\!\!  C=\frac{5\pi}{486m_{Q}^{4}}\langle \mathcal{O}(^{1}P_{1}^{[1]})\rangle_{^{1}P_{1}},
D=\frac{\pi}{3645m_{Q}^{6}}\langle \mathcal{P}(^{1}P_{1}^{[1]})\rangle_{^{1}P_{1}}. \nonumber
\end{eqnarray}

After the PMC scale setting, the PMC scale may be close to or even smaller than $\Lambda_{\rm QCD}$ in certain processes, which could lead to Landau pole problem for the running coupling. A small scale is reasonable since at higher orders more gluons are involved and all of them can share the typical momentum flow of the process and result in smaller renormalization scales. Such a small scale can explain the discrepancies between the conventional QCD predictions with the experimental data, e.g. it can shrink the gap between the pQCD estimation and experimental measurement for the top pair forward and backward asymmetry at the TEVATRON to be within $1\sigma$$^{[6]}$. Moreover, the commensurate scale relations among different renormalization schemes can smear such problem to a certain degree, since those relations between observables can be tested at quite low momentum transfers$^{[11]}$.

For the present $\eta_c$ and $h_{c}$ decay, their PMC scales are smaller than $1$ GeV. At the low energy region, the natural extension of the running coupling is somewhat questionable. To achieve a more accurate pQCD estimation, we adopt several low-energy models$^{[12-17]}$ suggested in the literature to do our discussion, i.e.
\begin{itemize}
\item The APT model$^{[12]}$, which is based on the analytic perturbative theory and takes the form
    \begin{equation}
    \alpha_{\rm APT}(\mu_R^{2}) =\frac{4\pi}{\beta_{0}}\left(\frac{1} {\ln x} +\frac{1}{1-x}\right),     \label{APT}
    \end{equation}
    where $x={\mu_R^{2}}/{\Lambda^{2}_{QCD}}$.
\item The WEB model$^{[13]}$, which is suggested by Webber to suppress the power correction of APT model and takes the form
    \begin{equation}
   \alpha_{\rm WEB}(\mu_R^{2})=\frac{4\pi}{\beta_{0}}\left[\frac{1}{\ln x}+\frac{x+b}{(1-x)(1+b)} \left(\frac{1+c}{x+c}\right)^p\right],
    \end{equation}
    where $b=1/4$ and $p=c=4$.
\end{itemize}
\end{multicols}
\begin{itemize}
\item The MPT model$^{[14]}$, which is based on the `massive' analytic pQCD theory. It suggests to use an effective glueball mass as the infrared regulator. The main idea of MPT is to change the logarithm $\ln \mu_R^2/\Lambda_{\rm QCD}^{2}$ by $\ln(\xi+\mu_R^2/\Lambda_{\rm QCD}^{2})$, in which $\xi$ corresponds to the ``effective gluonic mass" $m_{gl}=\sqrt{\xi}\Lambda_{\rm QCD}$. It takes the following form
    \begin{equation}
    \alpha_{\rm MPT}(\mu_R^{2})=\frac{\alpha_{\rm crit}}{1+\alpha_{\rm crit}\frac{\beta_{0}}{4\pi}\ln(1+x/\xi)+\alpha_{\rm crit}\frac{\beta_{1}}{2\pi\beta_{0}}\ln\left[1+\alpha_{\rm crit}\frac
    {\beta_{0}}{4\pi}\ln(1+x/\xi)\right]} ,
    \end{equation}
    where $\beta_{1}=51-19n_{f}/3$, $\alpha_{\rm crit}=0.61$ and $\xi=10$. It is noted that the moment of the spin-dependent structure function calculated within MPT model is consistent with the experiment data down to a few hundred MeV.
\end{itemize}
\begin{multicols}{2}
\begin{itemize}
\item The BPT model$^{[15]}$, which takes the form
    \begin{equation}
    \alpha_{\rm BPT}(\mu_R^{2}) =\frac{4\pi} {\beta_{0}t_{B}} \left(1-\frac{2\beta_{1}} {\beta_{0}^{2}}\frac{\ln t_{B}}{t_{B}}\right),
    \end{equation}
    where $t_{B}=\ln \frac{\mu_R^{2}+m_{B}^{2}}{\Lambda^{2}_{\rm QCD}}$ and $m_B=1$ GeV.
\item The CON model$^{[16]}$, which takes the form
    \begin{equation}
    \alpha_{\rm CON}(\mu_R^{2})=\frac{{4\pi}/{\beta_{0}}} {\ln\left[x+4M_{g}^{2}(\mu_R^{2})/\Lambda^{2}_{\rm QCD}\right]},
    \end{equation}
    where $M_{g}^{2}(\mu_R^{2})$ stands for the running gluon mass, determined by the gluon mass $m_{g}=0.34$:
    \begin{equation}
    M^2_{g}(\mu_R^{2})=m_{g}^{2}\left[\frac{\ln(x+4m_{g}^{2}/\Lambda^{2}_{\rm QCD})}{\ln(4m_{g}^{2}/\Lambda^{2}_{\rm QCD})}\right]^{-\frac{12}{11}}.
    \end{equation}
\item The GI model$^{[17]}$, which takes the form
    \begin{equation}
    \alpha_{\rm GI}(\mu_R^{2})=\sum^{3}_{k=1}\alpha_{k}\rm exp\left[-\mu_R^{2}/4\gamma_{k}^{2}\right],
    \end{equation}
    where $\alpha_{1}=0.25$, $\alpha_{2}=0.15$, $\alpha_{3}=0.2$, $\gamma_{1}^{2}=1/4$, $\gamma_{2}^{2}=5/2$ and $\gamma_{3}^{2}=250$.
\end{itemize}

\vskip 4mm

\centerline{\includegraphics[width=0.5\textwidth]{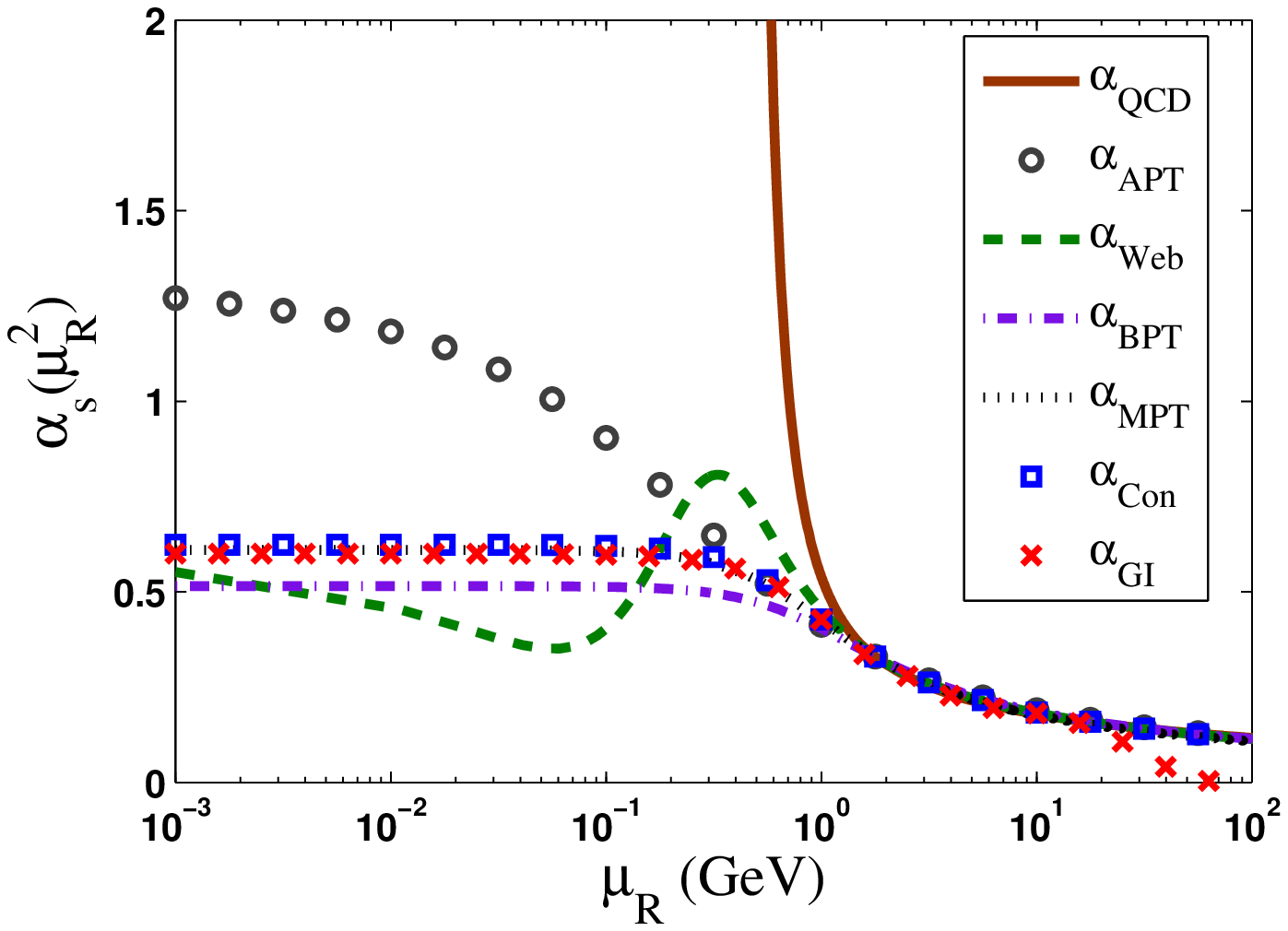}}

\vskip 2mm

\centerline{\footnotesize \begin{tabular}{p{7.5 cm}}\bf Fig.\,1. \rm
Various effective running strong coupling models versus the scale $\mu_R$. $\alpha_{\rm QCD}$ stands for the conventional behavior for the strong running coupling.
\end{tabular}}

\medskip

Since the behavior of the running coupling is universal, the above model parameters have been determined by each group via comparing with the known data. Further more, to apply those low-energy models, we also need to set the scale parameter $\Lambda_{\rm QCD}$. Using the two-loop $\alpha_s$ running with $\alpha(\rm M_{\tau})=0.33$ $^{[18]}$, we can predict $\Lambda_{\rm QCD}$ for the conventional QCD running behavior, i.e. we have $\Lambda^{(n_f=3)}_{\rm QCD}=0.387$ GeV, $\Lambda^{(n_f=4)}_{\rm QCD}=0.333$ GeV and $\Lambda^{(n_f=5)}_{\rm QCD}=0.231$ GeV. For APT model, we have $\Lambda^{(n_f=3)}_{\rm QCD}=0.254$ GeV. For WEB model, we have $\Lambda^{(n_f=3)}_{\rm QCD}=0.214$ GeV. For BPT model, we have $\Lambda^{(n_f=3)}_{\rm QCD}=0.453$ GeV. For MPT model, we have $\Lambda^{(n_f=3)}_{\rm QCD}=0.294$ GeV. For CON model, we have $\Lambda^{(n_f=3)}_{\rm QCD}=0.222$ GeV. A comparison of running coupling has been presented in Fig.1. Except for the GI model, Fig.1.shows that as required, the low-energy models mainly change the low-energy behavior and their high-energy behaviors are almost unchanged in comparison with the conventional running coupling.

For the color-singlet LDMEs, they can be related to the wavefunction at the origin for the $S$-wave states or the first derivative of the wavefunction at the origin for the $P$-wave states. We adopt their values derived from the B-T potential model$^{[19]}$ to fix those LDMEs, i.e. $|R_{\eta_{c}}|^{2}=0.810{\rm GeV^{3}}$, $|R_{\eta_{b}}|^{2}=6.477{\rm GeV^{3}}$, $|R'_{h_{c}}|^{2}=0.075{\rm GeV^{5}}$, $|R'_{h_{b}}|^{2}=1.417{\rm GeV^{5}}$. For the charmonium LDMEs involving $S$-wave state, we obtain $\langle{\mathcal{O}(^{1}S^{[1]}_{0})}\rangle_{\eta_{c}}=0.387{\rm GeV^{3}}$ and $\langle{\mathcal{P}(^{1}S^{[1]}_{0})}\rangle_{\eta_{c}} =m_{c}^{2}\langle\upsilon^{2}\rangle_{\eta_{c}}\langle
{\mathcal{O}(^{1}S^{[1]}_{0})}\rangle_{\eta_{c}}=0.198{\rm GeV^{5}}$. For the LDMEs involving $P$-wave state, we have $\langle{\mathcal{O}(^{1}P^{[1]}_{1})}\rangle_{h_c}= 0.107{\rm GeV^{5}}$ and $\langle{\mathcal{P}(^{1}P^{[1]}_{1})} \rangle_{h_c}= m_{c}^{2}\langle \upsilon^{2}\rangle_{h_c}\langle {\mathcal{O}(^{1}P^{[1]}_{1})} \rangle_{h_c}=0.055{\rm GeV^{7}}$. The color-octet LDMEs can be determined by the evolution equations$^{[2,3,20,21]}$:
\begin{eqnarray}
&& \!\!\!\!\!\!\!\!\!\!\! \mu_{F}^{2}\frac{d\langle{\mathcal{O}(^{1}S^{[8]}_{0})}\rangle} {d\mu_{F}^{2}}
=-\frac{7\alpha_{s}}{9\pi}\frac{\langle{\mathcal{P}(^{1}S^{[8]}_{0})} \rangle}{m_{c}^{2}}+\frac{16
\alpha_{s}}{9\pi}\frac{\langle{\mathcal{O}(^{1}P^{[1]}_{1})}\rangle} {2N_{c}m_{c}^{2}}, \nonumber \\
&&\quad\quad\quad\quad\quad\quad -\frac{16\alpha_{s}}{15\pi}\frac{\langle{\mathcal{P}(^{1}P^{[1]}_{1})}\rangle} {2N_{c}m_{c}^{4}},  \\
&& \!\!\!\!\!\!\!\!\!\!\! \mu_{F}^{2}\frac{d\langle{\mathcal{P}(^{1}S^{[8]}_{0})}\rangle} {d\mu_{F}^{2}}
=\frac{16\alpha_{s}}{9\pi}\frac{\langle{\mathcal{P}(^{1}P^{[1]}_{1})}\rangle} {2N_{c}m_{c}^{2}}.
\end{eqnarray}
Setting $\mu_F=2m_c$, we obtain $\langle{\mathcal{O}(^{1}S^{[8]}_{0})} \rangle_{h_c}=0.0045{\rm GeV^{3}}$ and $\langle{\mathcal{P}(^{1}S^{[8]}_{0})} \rangle_{h_c}=0.0028{\rm GeV^{5}}$. For the case of bottomonium, we have $\langle{\mathcal{O}(^{1}S^{[1]}_{0})} \rangle_{\eta_{b}}=3.092{\rm GeV^{3}}$, $\langle{\mathcal{P}(^{1}S^{[1]}_{0})} \rangle_{\eta_{b}}=2.748{\rm GeV^{5}}$, $\langle{\mathcal{O}(^{1}P^{[1]}_{1})} \rangle_{h_b}=2.030{\rm GeV^{5}}$, $\langle{\mathcal{P}(^{1}P^{[1]}_{1})} \rangle_{h_b}=1.804{\rm GeV^{7}}$, $\langle{\mathcal{O}(^{1}S^{[8]}_{0})} \rangle_{h_b}=0.0074{\rm GeV^{3}}$ and $\langle{\mathcal{P}(^{1}S^{[8]}_{0})} \rangle_{h_b}=0.0068{\rm GeV^{5}}$. In the calculation, we have taken $\langle\upsilon^{2} \rangle_{h_{c}} \approx\langle\upsilon^{2}\rangle_{\eta_{c}}=0.228$$^{[1]}$ and $\langle\upsilon^{2}\rangle_{h_{b}}\approx \langle\upsilon^{2} \rangle_{\eta_{b}} =0.042$$^{[2]}$. When varying the quark masses, those LDMEs shall be changed accordingly.

\end{multicols}

\vskip 2mm

\noindent{\footnotesize \textbf{Table 1}. Total decay widths for $\Gamma(H[n])$  under the PMC scale setting with various low-energy running coupling models, where $\mu^{\rm init}_{R}=2m_{Q}$ and $[n]=^1S^{(1)}_0$ or $^1P^{(1)}_1$. The errors are calculated by taking $m_{c}\in[1.40\rm GeV,1.60\rm GeV]$ and $m_{b}\in[4.50\rm GeV,4.70\rm GeV]$.

\vskip 2mm \tabcolsep 10pt

\centerline{\footnotesize
\begin{tabular}{cccccccc}\hline
$  $ & QCD & WEB & APT & BPT & MPT& CON & GI \\
\hline $\Gamma(\eta_{c})$ (MeV)
 & $41.70^{+16.17}_{-10.45}$ & $28.63^{+7.43}_{-5.50}$ & $23.92^{+5.31}_{-4.07}$ & $26.70^{+6.15}_{-4.73}$ & $25.09^{+5.52}_{-4.28}$ & $25.41^{+5.73}_{-4.42}$ & $25.59^{+5.87}_{-4.55}$ \\
$\Gamma(\eta_{b})$ (MeV)
 & $13.80^{+0.89}_{-0.82}$ & $14.33^{+0.90}_{-0.83}$ & $14.97^{+0.91}_{-0.84}$ & $14.08^{+0.91}_{-0.84}$ & $14.34^{+0.92}_{-0.84}$ & $14.54^{+0.91}_{-0.84}$ & $13.88^{+0.86}_{-0.79}$  \\
$\Gamma(h_{c})$ (MeV)
 & $0.87^{+0.12}_{-0.11}$ & $0.61^{+0.07}_{-0.05}$ & $0.51^{+0.06}_{-0.04}$ & $0.58^{+0.06}_{-0.05}$ & $0.54^{+0.06}_{-0.04}$ & $0.55^{+0.05}_{-0.05}$ & $0.55^{+0.06}_{-0.05}$  \\
$\Gamma(h_{b})$ (KeV)
 & $38.88^{+0.38}_{-0.54}$ & $39.86^{+0.27}_{-0.45}$ & $41.01^{+0.14}_{-0.33}$ & $39.40^{+0.34}_{-0.51}$ & $39.89^{+0.28}_{-0.46}$ & $40.24^{+0.23}_{-0.41}$ & $39.04^{+0.30}_{-0.47}$   \\
\hline
\end{tabular}}}

\begin{multicols}{2}

After applying the PMC scale setting, we present the total decay widths for the conventional and the six low-energy effective running coupling models in Table 1. As for the bottomonium case, the PMC scales are larger than $1$ GeV, the results of all low-energy models agree well with each other. As for the charmonium case, the conventional running coupling could be questionable, since the PMC scales for $\eta_{c}$ and $h_c$ are less than $1$ GeV, i.e. $\mu^{\rm PMC}_{\eta_{c}}=0.93$ GeV and $\mu^{\rm PMC}_{h_{c}}=0.98$ GeV. It is found that the total decay widths with the six low-energy effective models are consistent with each other, which are caused by the fact that those models have similar behavior around the region close to $1$ GeV. In the following, we adopt MPT model to do a detail discussion on the heavy quarkonium decay properties.

\vskip 2mm

\noindent{\footnotesize \textbf{Table 2}. Scale dependence of $\Gamma(H[n])$ within the MPT model under the conventional scale setting and the PMC scale setting, where three typical (initial) scales are adopted.

\vskip 2mm \tabcolsep 4pt

\centerline{\footnotesize
\begin{tabular}{cccccccc}\hline
 & \multicolumn{3}{c}{Conventional }  &  \multicolumn{3}{c}{~~~~~~~~~PMC }\\
\cline{2-4} \cline{6-8}
$\mu^{\rm init}_{R}$ & $\rm m_{Q}$ & $2\rm m_{Q}$ & $4\rm m_{Q}$ && $\rm m_{Q}$ & ${2\rm m_{Q}}$ & $4\rm m_{Q}$ \\
\hline $\Gamma(\eta_{c})$ (MeV)
 & 27.78 & 20.38 & 14.01 && 25.09 & 25.09 & 25.09   \\
$\Gamma(\eta_{b})$ (MeV)
 & 12.91 & 10.14 & 7.94 && 14.34 & 14.34 & 14.34  \\
$\Gamma(h_{c})$ (MeV)
 & 0.57 & 0.41 & 0.28 && 0.54 & 0.54 & 0.54  \\
$\Gamma(h_{b})$ (KeV)
 & 45.92 & 39.27 & 31.98 && 39.89 & 39.89 & 39.89  \\
\hline
\end{tabular}}}

\vskip 0.5\baselineskip

\vskip 2mm

\noindent{\footnotesize \textbf{Table 3}. Total decay width $\Gamma(H[n])$ within the MPT model under the conventional scale setting and the PMC scale setting, where $\mu^{\rm init}_R=2\rm m_{Q}$.

\vskip 2mm \tabcolsep 4pt

\centerline{\footnotesize
\begin{tabular}{cccccccc}\hline
 & \multicolumn{3}{c}{Conventional }  &  \multicolumn{3}{c}{~~~~~~~~~PMC }\\
\cline{2-4} \cline{6-8}
$  $ & ~LO~ & ~NLO~ & ~sum~ && ~LO~& ~NLO~& ~sum~  \\
\hline $\eta_{c}$ (MeV)
 & 11.62 & 8.76 & 20.38 && 31.95 & -6.85 & 25.09  \\
$\eta_{b}$ (MeV)
 & 6.25 & 3.89 & 10.14 && 15.68 & -1.33 & 14.35 \\
$h_{c}$ (MeV)
& 0.23 & 0.18 & 0.41 && 0.61 & -0.07 & 0.54 \\
$h_{b}$ (KeV)
 & 27.97 & 11.30 & 39.27 && 70.12 & -30.23 & 39.89 \\
\hline
\end{tabular}}}

\vskip 0.5\baselineskip

We present the decay widths under the MPT model before and after the PMC scale setting in Table 2 and Table 3. As shown by Table 2, under the conventional scale setting, the decay widths depend heavily on the choice of scale. By varying $\mu_R\equiv\mu^{\rm init}_R$ from $m_Q$ to $4m_Q$, the decay widths for $\eta_{c}$, $\eta_{b}$, $h_c$ and $h_b$ are changed by about $50\%$, $39\%$, $51\%$ and $30\%$ respectively. On the other hand, after the PMC scale setting, we observe that all the decay widths remain almost unchanged, thus the scale ambiguity is eliminated even at the NLO level. There is residual scale dependence due to unknown higher-order $\{\beta_i\}$-terms, which however will be highly exponentially suppressed$^{[5]}$. After the PMC scale setting, one can absorb/resum the $\{\beta_{0}\}$-terms into the running coupling, and in principle, the LO decay widths shall be increased and the NLO decay widths shall be decreased. Thus, the pQCD convergence can be generally improved. As shown by Table 3, this is indeed the case for most of the decays. For convenience, we define a parameter $K$, which equals to the ratio of the NLO decay width to the LO decay width. Under the conventional scale setting, the $K$ factors for $\eta_{c}$, $\eta_{b}$ and $h_c$ and $h_b$ are $75\%$, $62\%$ and $78\%$, which are changed down to $21\%$, $8\%$ and $11\%$ after the PMC scale setting. The only exception is the $h_b$ decay, whose $K$ factor is about $40\%$ even after PMC scale setting, which means one needs to finish at least the NNLO calculation to achieve a better pQCD convergence for the $h_b$ decay.

In the literature, another scale setting method based on the local renormalization group invariance, i.e. the principle of minimum sensitivity (PMS)$^{[22]}$, has also been suggested. Following the standard PMS procedures, we obtain
\begin{eqnarray}
\Gamma(H[n])=C_{0}([n])\alpha_{s}^{2}(\mu_{R}^{\rm PMS}) \frac{3+c\alpha_{s}(\mu_{R}^{\rm PMS})/\pi}{3\left(1+c\alpha_{s}(\mu_{R}^{\rm PMS})/\pi\right)},
\end{eqnarray}
where $c=\beta_{1}/2\beta_{0}$, and the PMS effective coupling is a solution of the following equation:
\begin{eqnarray}
&& \!\!\!\!\!\!\!\!\!\!\! \rho_{1}([n])=\frac{2\pi}{\alpha_{s}(\mu_{R}^{\rm PMS})}+2c\ln\left(\frac{c\alpha_{s}(\mu_{R}^{\rm PMS})/\pi}{1+c\alpha_{s}(\mu_{R}^{\rm PMS})/\pi}\right) \nonumber \\
&&\quad\quad+\frac{2c}{3(1+c\alpha_{s}(\mu_{R}^{\rm PMS})/\pi)}.
\end{eqnarray}
Here $\rho_{1}([n])=\frac{\beta_{0}}{2}\ln\left(\frac{\mu_{R}^{2}(\rm init)}{\widetilde{\Lambda}^{2}_{\rm QCD}}\right)-C_{1}([n])$ and $\widetilde{\Lambda}_{\rm QCD} =\Lambda_{\rm QCD}\left(\frac{2\beta_{1}} {\beta_{0}^{2}} \right)^{-\beta_{1}/\beta_{0}^{2}}$.

\vskip 2mm

\noindent{\footnotesize \textbf{Table 4}. Comparison of PMC and PMS estimations for $\Gamma(H[n])$ together with the experimental measurements$^{[18]}$. The MPT model is adopted and the errors are calculated by taking $m_{c}\in[1.40\rm GeV,1.60\rm GeV]$ and $m_{b}\in[4.50\rm GeV,4.70\rm GeV]$.

\vskip 2mm \tabcolsep 4pt

\centerline{\footnotesize
\begin{tabular}{cccc}\hline
&~~~experiment~~~ &~~~PMC~~~&~~~PMS\\
\hline $\Gamma(\eta_{c})$ (MeV)
& $32.0\pm0.9$ & $25.09^{+5.52}_{-4.28}$ & $31.57^{+9.06}_{-6.47}$\\
$\Gamma(\eta_{b})$ (MeV)
& $10.8^{+4.0+4.5}_{-3.7-2.0}$ & $14.34^{+0.92}_{-0.84}$ &$13.25^{+0.81}_{-0.75}$ \\
$\Gamma(h_{c})$ (MeV)
& $\sim$ & $0.54^{+0.06}_{-0.04}$ & $0.66^{+0.08}_{-0.06}$ \\
$\Gamma(h_{b})$ (KeV)
& $\sim$ & $39.89^{+0.28}_{-0.46}$ & $42.95^{+1.42}_{-1.38}$ \\
\hline
\end{tabular}}}

\vskip 0.5\baselineskip

We present a comparison of the PMC and PMS estimations by using the MPT model in Table 4, in which the available experimental results are also presented. Due to large errors from the bound state parameters such as the quark masses, both the PMC and PMS estimations are consistent with the experimental results within reasonable regions. As an estimation of $h_c$ decay, it has two dominant decay channels, by taking $\Gamma(h_c\to\gamma\eta_{c})=385$ KeV$^{[23]}$, we obtain $\Gamma^{th}(h_c)\simeq 0.925$ MeV, which is consistent with $\Gamma^{\rm exp}(h_c)=0.73\pm0.45\pm0.28$ MeV$^{[24]}$. As for $h_b$, its decay width $\Gamma(\eta_{b})$ has been estimated by Refs.[25,26,27]. By taking $\Gamma(h_b\to \gamma+\eta_{b})=37.0$ KeV$^{[26]}$, we obtain ${\cal B}^{th}(h_b(1P) \to\eta_{b}(1S)\gamma) =\left(48.1^{+0.3}_{-0.2} \right)\%$, which agrees ${\cal B}^{exp}(h_b(1P)\rightarrow\eta_{b}(1S)\gamma) = \left(49.2\pm5.7^{+5.6}_{-3.3}\right)\%$$^{[18]}$.

In summary, we have made a detailed discussion on the decay widths for the spin-singlet heavy quarkoniums under the PMC scale setting. After the PMC scale setting, the renormalization scale uncertainty has been eliminated even at the NLO level. Thus, after applying the PMC scale setting, it can eliminate an important theoretical error and increase the precision of QCD tests, which shall also increase the sensitivity of the collider experiments to new physics beyond the standard model$^{[28]}$. Moreover, we show that the PMC estimations are also consistent with the PMS estimations for the present decay channels. The remaining uncertainties are from the bound state parameters such as the constituent $b$ or $c$ quark mass. As a final remark, it is noted that the PMC scales of $\eta_{c}$ and $h_c$ are below $1$ GeV, in such low-energy region, a proper low-energy running coupling is necessary. We have applied several low-energy running coupling models for the estimation. The results show that the decay widths under those models have similar results and are in reasonably consistent with the data.

\section*{\Large\bf References}

\vspace*{-0.8\baselineskip}

\hskip 7pt {\footnotesize

\REF{[1]} Li J Z, Ma Y Q and Chao K T 2011 {\it Phys. Rev.} D {\bf 83} 114038.

\REF{[2]} Li J Z, Ma Y Q and Chao K T 2013 {\it Phys. Rev.} D {\bf 88} 034002.

\REF{[3]} Bodwin G T, Braaten E and Lepage G P 1995 {\it Phys. Rev.} D {\bf 51}, 1125 ; 1997 Erratum-ibid. D {\bf 55} 5853.

\REF{[4]} Brodsky S J and Wu X G 2012 {\it Phys. Rev. Lett.} {\bf 109} 042002.

\REF{[5]} Brodsky S J and Wu X G 2012 {\it Phys. Rev.} D {\bf 85} 034038.

\REF{[6]} Brodsky S J and Wu X G 2012 {\it Phys. Rev.} D {\bf 85} 114040.

\REF{[7]} Brodsky S J and Wu X G 2012 {\it Phys. Rev.} D {\bf 86} 014021.

\REF{[8]} Brodsky S J and Wu X G 2012 {\it Phys. Rev.} D {\bf 86} 054018.

\REF{[9]} Mojaza M, Brodsky S J and Wu X G 2013 {\it Phys. Rev. Lett.} {\bf 110} 192001; Brodsky S J, Mojaza M and Wu X G 2014 {\it Phys. Rev.} D {\bf 89} 014027.

\REF{[10]} Brodsky S J and Giustino L D 2012 {\it Phys. Rev.} D {\bf 86} 085026.

\REF{[11]} Brodsky S J and Lu H J 1995 {\it Phys. Rev.} D {\bf 51} 3652.

\REF{[12]} Shirkov D V and Solovtsov I L 1997 {\it Phys. Rev. Lett.} {\bf 79} 1209.

\REF{[13]} Webber B R 1998 {\it JHEP} {\bf 10} 012.

\REF{[14]} Shirkov D V 2013 {\it Phys. Part. Nucl. Lett.} {\bf 10} 186.

\REF{[15]} Badalian A M and Kuzmenko D S 2002 {\it Phys. Rev.} D {\bf 65} 016004.

\REF{[16]} Cornwall J M 1982 {\it Phys. Rev.} D {\bf 26} 1453.

\REF{[17]} Godfrey S and Isgur N 1985 {\it Phys. Rev.} D {\bf 32} 189.

\REF{[18]} Beringer J et al [Particle Data Group] 2012 {\it Phys. Rev.} D {\bf 86} 010001.

\REF{[19]}  Eichten E J and Quigg C 1995 {\it Phys. Rev.} D {\bf 52} 1726.

\REF{[20]} Gremm M and Kapustin A 1997 {\it Phys. Lett.} B {\bf 407} 323.

\REF{[21]} Fan Y, Li J Z, Meng C and Chao K T 2012 {\it Phys. Rev.} D {\bf 85} 034032.

\REF{[22]} Stevenson P M 1981 {\it Phys. Rev.} D {\bf 23} 2916.

\REF{[23]} Chao K T, Ding Y B and Qin D H 1993 {\it Phys. Lett.} B {\bf 301} 282.

\REF{[24]} Ablikim M et al [The BESIII Collaboration] 2010 {\it Phys. Rev. Lett.} {\bf 104} 132002.

\REF{[25]} Brambilla N et al [Quarkonium Working Group], hep-ph/0412158v2.

\REF{[26]} Godfrey S and Rosner J L 2002 {\it Phys. Rev.} D {\bf 66} 014012.

\REF{[27]} Li B Q and Chao K T 2009 {\it Commun. Theor. Phys.} {\bf 52} 653.

\REF{[28]} Wu X G, Brodsky S J and Mojaza M 2013 {\it Prog. Part. Nucl. Phys.} {\bf 72} 44.

}

\end{multicols}

\end{document}